\documentclass[11pt,aps,prd,preprint,superscriptaddress]{revtex4-1}
\usepackage{amsmath}
\usepackage[dvips]{graphicx}
\usepackage[english]{babel}
\newcommand{\be}{\begin{equation}}
\newcommand{\ee}{\end{equation}}
\newcommand{\ben}{\begin{eqnarray}}
\newcommand{\een}{\end{eqnarray}}

\begin{document}
\title{On the $c^{2}$ term  in the holographic formula for dark energy}
\author{Ninfa Radicella\footnote{E-mail: ninfa.radicella@uab.cat} and
Diego Pav\'{o}n\footnote{E-mail: diego.pavon@uab.es}}
\affiliation{Departamento de F\'{\i}sica, Universidad Aut\'{o}noma
de Barcelona, 08193 Bellaterra (Barcelona), Spain.}
\begin{abstract}
It is argued that the $c^{2}$ term that appears in the
conventional formula for holographic dark energy should not be
assumed constant in general. Notwithstanding, there is at least an
exception, namely, when the Ricci scale is chosen as the infrared
cutoff length.
\end{abstract}
\maketitle
\section{Introduction}
Cosmological models based on holographic dark energy  rest on the
rather reasonable assumption that the entropy of every bounded
region of the Universe, of size $L$, should not exceed the entropy
of a Schwarzschild black hole of the same size. Mathematically,
\be L^{3}\, \Lambda^{3} \leq S_{BH} \simeq L^{2}M_{Pl}^{2}\,
\quad\qquad (M_{Pl}^{2} = (8\pi G)^{-1})\, ,
 \label{bound1}
\ee
where $\Lambda$ stands for the ultraviolet cutoff while the
infrared cutoff is set by $L$.

However, an effective field theory that saturates the above
inequality necessarily includes states such that the Schwarzschild
radius exceeds $L$ \cite{cohen}. It is therefore natural to
replace the above bound by another  not allowing such states,
\be L^{3}\, \Lambda^{4} \leq M_{Pl}^{2}\, L \, .  \label{bound2}
\ee
This bound guarantees that the energy $L^{3}\Lambda^{4}$ in a
region of the size $L$ does not exceed the energy of a black hole
of the same size \cite{mli1}. By saturating the inequality
(\ref{bound2}) and identifying $\Lambda^{4}$ with the density of
holographic dark energy, $\rho_{x}$, it follows that
\cite{cohen,mli1}
\be \rho_{x}= \frac{3 c^{2}}{8\pi G \, L^{2}}\, , \label{rhox} \ee
where the factor $3$ was introduced for convenience and $c^{2}$ is
a dimensionless quantity that collects the uncertainties of the
theory (such as the number particle species and so on). On the
other hand, a further interesting feature of holography lies in
its close connection with the spacetime foam, as unveiled by Ng
\cite{jng, arzano}. Additional motivations for holographic dark
energy can be found in Section 3 of \cite{wd-cqg}.

Very often, for the sake of simplicity, the $c^{2}$ parameter is
assumed constant. However, one should bear in mind that it is more
general to consider it a slowly varying function of time,
$c^{2}(t)$, as in \cite{dw-plb, d-jphys, raul}. By slowly varying
we mean that $(c^{2})^{\cdot}/c^{2}$ is upper bounded by the
Hubble expansion rate, $H$, i.e.,
\be \frac{(c^{2}(t))^{\cdot}}{c^{2}(t)} \lesssim H  \, .
\label{condition} \ee
Note that this condition must be fulfilled at all times; otherwise
the dark energy density would not even approximately be
proportional to $L^{-2}$, something at the core of holography.

The target of this work is to argue that, rather generally, it is
not consistent to consider $c^{2}$ constant. Obviously, $c^{2}$
will depend on the infrared length, $L$ assumed by the model. We
shall consider three different lengths: the Hubble length,
$H^{-1}$, the particle horizon length (defined below by Eq.
(\ref{parthor})), and Ricci's length, $L = (\dot{H}\, + 2
H^{2})^{-1/2}$. We shall not include in our discussion the widely
used length defined by the radius of the future event horizon
because the corresponding models suffer from a severe circularity
problem. Namely, the event horizon is needed to have acceleration
and vice versa.

As for dark energy, we shall consider first the cosmological
constant and then dark energy with arbitrary equation of state. In
all the cases, we shall limit ourselves to spatially flat
universes described by the Friedmann-Robertson-Walker (FRW)
metric.

\section{Dark energy given by the cosmological constant}
Nowadays the spatially flat $\Lambda$CDM model is, from the
observational point of view, the leading cosmological model in the
market -see, e.g. \cite{ann-rev, Amendola, serra09, li-li}. It
assumes that the energy budget is dominated by the constant dark
energy density of the quantum vacuum, $\rho_{\Lambda}$, and a dust
contribution that redshifts with expansion quickly while radiation
and other forms of energy are negligible at present, which also
redshift with expansion. This implies that $\rho_{\Lambda}
\rightarrow \rho_{total}$ as $t \rightarrow \infty$.

Before going any further one may ask, in the first place, why
$\rho_{\Lambda}$ should be holographic at all. Recall that the
entropy of the quantum vacuum is a constant that may be fixed to
zero. We have no clear answer to this point. However, if dark
energy with equation of state $w = - 1 + \epsilon$ (with
$|\epsilon| \ll 1$) is holographic, then $\rho_{\Lambda}$ should
also be holographic, on account of continuity, when $\epsilon
\rightarrow 0$. Thus, we shall tentatively assume that
$\rho_{\Lambda}$ obeys Eq. (\ref{rhox}).

To motivate that $c^{2}$ must vary with time we take first the
infrared cutoff provided by the Hubble length, $L = H^{-1}$.
Barring the presence of fields that violate the dominant energy
condition, we can replace $\rho_{total}$ by $\rho_{\Lambda}$ at
very late times, whence in virtue of the first Friedmann equation,
we can write
\be
\rho_{\Lambda} = \frac{3}{8\pi G}\, c^{2}(t \rightarrow
\infty) \, H^{2}(t \rightarrow \infty) = \frac{3}{8\pi G} \,
H^{2}(t \rightarrow \infty) \, \qquad  \Rightarrow \qquad  c^{2}(t
\rightarrow \infty) := c^{2}_{\infty} = 1 \, .
\label{latetimes}
\ee

\noindent Obviously, $c^{2}$ cannot be unity at earlier times
because it would not leave room for any other forms of energy
whatsoever (matter, radiation, ...). On the other hand it is
obvious that $c^{2}$ approaches unity from below.

Another point to consider is whether $c^{2}$ varies sufficiently
slow at all times, i.e., whether the condition (\ref{condition})
is met at all epochs.  To elucidate this we recall that
$\rho_{\Lambda} =$ constant. Then,
\be
c^{2}(t)\, H^{2}(t) = c^{2}_{\infty}\, H^{2}_{\infty} = {\rm
constant} \qquad \Rightarrow \qquad [c^{2}\, H^{2}]^{\cdot} = 0 \,
. \label{c2H2}
\ee
To make matters easier, assume that the Universe is dominated
solely by pressureless matter (subscript $m$) and
$\rho_{\Lambda}$. Using the second Friedmann equation, $\dot{H} =
-4 \pi G \rho_{m}$, and $\Omega_{m} := 8 \pi G \rho_{m}/(3H^{2})$
(bear in mind that $\Omega_{m}$ varies with time), we get
\be
\frac{(c^{2})^{\cdot}}{c^{2}} = 3 \, \Omega_{m}\, H \, .
\label{c2dot1}
\ee
At early times, $\Omega_{m} \simeq 1$ and condition
(\ref{condition}) is violated. We conclude by saying that if $ L =
H^{-1}$, then the energy density of the quantum vacuum is not
holographic at early times; it might be thought as holographic
only when $\Omega_{m} \lesssim 1/3$.

\bigskip

Assume now that the infrared cutoff is given by Ricci's length, $L
= (\dot{H}\, + 2 H^{2})^{-1/2}$, as in \cite{gao}.
Since $\rho_{\Lambda}$ is to completely dominate over
non-relativistic matter when $t \rightarrow \infty$ it follows
that
\be
\rho_{\Lambda} = \frac{3H^{2}_{\infty}}{8 \pi G} = {\rm
constant} = \frac{3 c^{2}}{8 \pi G} \, (\dot{H}(t) \, + \, 2
H^{2}(t)) \, , \label{latetimesagain} \ee
but $\dot{H}(t \rightarrow \infty)$ vanishes because $\dot{H} =
-4\pi G \rho_{m}$. As a consequence, $c^{2}_{\infty} = 1/2$. This
value is attained asymptotically from below.

Thus, as expected, the asymptotic value of $c^{2}$ depends on the
infrared cutoff.

To elucidate whether condition (\ref{condition}) is fulfilled, we
proceed as before. We start from $[c^{2}\, (\dot{H} \, + \,
2H^{2})]^{\cdot} = 0$ and use $\ddot{H} = -4 \pi G \dot{\rho}_{m}
= 12 \pi G H \rho_{m}$  alongside the expressions for $\dot{H}$
and $\Omega_{m}$ of above to obtain
\be \frac{(c^{2})^{\cdot}}{c^{2}} = \frac{3 \, \Omega_{m}}{4\, -\,
3 \Omega_{m}}H  \, . \label{c2dot2} \ee
In consequence, when $2/3 \lesssim \Omega_{m} \lesssim 1$ (i.e.,
early times) one has $H \lesssim (c^{2})^{\cdot}/c^2$. So, in this
case as well $\rho_{\Lambda}$ fails to be holographic.

\bigskip

We next take the particle horizon,
\be L=a(t)\int_0^t\frac{dt'}{a(t')}=a\int_0^a\frac{da'}{a'{^2}
H(a')}, \label{parthor} \ee
where $a$ is the scale factor of the FRW metric, as the infrared
cutoff. For $\rho_{\Lambda}$ to be consistent with (\ref{rhox}),
$c(t)$ must vary as $L$.  \\

For a spatially flat FRW universe dominated by dust and a
cosmological constant, the first Friedmann's equation reduces to
\be H^2=\frac{8\pi G \rho_{\Lambda}}{3} \left[\frac{r_0}{a^3}+1\right],
\label{friedmattcosm}\ee
where $r_0$ denotes the present value of the ratio between the
densities of matter and vacuum energy,
$r\equiv\rho_m/\rho_\Lambda$. In this case $L$ can be expressed in
terms of a hypergeometric function, and we obtain
 \be c(a)=\sqrt{\frac{8\pi G \rho_{\Lambda}}{3}}L=
 2\sqrt{\frac{a^3}{r_0}}\  _2F_1\left(1/6,1/2,7/6;-a^3/r_0\right).
\label{cpart}
\ee
Next we  check whether for such a choice of the infrared cutoff
$c^2$ varies slowly enough. Condition (\ref{condition}) leads to
\be \frac{(c^2)^\cdot}{c^2}=\frac{(L^2)^\cdot}{L^2}=2
\left[H+\frac{1}{a^2 H\int_0^a\frac{da'}{a'^{2}
H(a')}}\right]\lesssim  H \, . \label{c2dot3}\ee
That is to say,
\be H+\frac{2}{\sqrt{\frac{a^5}{r_0}}H \ _2 F_1
\left(1/6,1/2,7/6;-a^3/r_0\right)}<0\, .
\label{ineq}\ee
Obviously the inequality fails since all the left hand side terms
are non-negative. In particular, the hypergeometric function, $_2
F_1 \left(1/6,1/2,7/6;-a^3/r_0\right)$, varies monotonously from
zero at $a = 0$ to $\; 2\, r_0^{1/3} \Gamma(1/3)
\Gamma(7/6)/\sqrt{\pi}\, $ when $\, a\rightarrow \infty$.

\section{Dark energy with generic equation of state}
Consider the Universe dominated by pressureless matter
 and dark energy (subindexes $x$ and $m$, respectively) and assume
 that they interact with each other gravitationally only. Then,
 \be
\rho_{m} = \rho_{m0}\, (1+z)^{3}\, , \quad {\rm and} \quad
\rho_{x} = \rho_{x0}\,
 \exp \left[3 \int_{0}^{z}{\left(\frac{1+w(z')}{1+z'} \right)dz'}\right]\,
 .
 \label{conserv}
 \ee
(We note in passing that, in general,  the equation of state
parameter of dark energy, $w$, is not constant in this context).
On the other hand, if the dark energy is holographic with the
infrared length given by the Hubble radius, $L = H^{-1}$, then
\be
\rho_{x} = \frac{3c^{2}}{8 \pi G}\, H^{2} \, .
\label{rhox-holographic}
\ee
Recalling the first Friedmann's equation, $3 H^{2} = 8 \pi G
(\rho_{m} \, + \, \rho_{x})$, we can write
\be \frac{3H^{2}}{8 \pi G}\, (1 \, - \, c^{2}) = \rho_{m0} \,
(1+z)^{3} \label{Friedmann1} \ee
as well as
\be
\frac{3H^{2}}{8 \pi G} = \rho_{m0}\, (1+z)^{3} \, + \, \rho_{x0}\,
 \exp \left[3 \int_{0}^{z}{\left(\frac{1+w(z')}{1+z'} \right)dz'}\right]\, .
\label{Friedmann2}
\ee
After combining last two equations and simplifying, we get
\be r = \frac{1 \, - \, c^{2}}{c^{2}} = r_{0} \, (1+z)^{3} \,\exp
\left[-3 \int_{0}^{z}{\left(\frac{1+w(z')}{1+z'}
\right)dz'}\right]  \, . \label{simplifying} \ee
That is to say, as the second equality tells us, the $c^{2}$
parameter cannot but vary with expansion. Moreover, as $z
\rightarrow -1$ (i.e., $t \rightarrow \infty$), $c^{2} \rightarrow
1$ from below and $r \rightarrow 0$.

We next study whether $c^{2}$ evolves slowly enough. To this end
we use the conservation equation $\dot{\rho_{x}} \, + \,
3H(1+w)\rho_{x} = 0$ alongside Eq. (\ref{rhox-holographic}) to
obtain
\[
\frac{(c^{2})^{\cdot}}{c^{2}} H \, + \, 2\dot{H} = - 3(1+w) H^{2}
\, .
\]
With the help of the first Friedmann equation (above) and the
second one, \newline $\dot{H} = - 4 \pi G \left[\rho_{m} \, +
\,(1+w)\rho_{x}\right]\,$, it follows that
$((c^{2})^{\cdot}/c^{2}) \, H  = -8 \pi G w \rho_{m}$. Dividing by
$H^{2}$  yields
\be \frac{(c^{2})^{\cdot}}{c^{2}} = - 3Hw\, \Omega_{m} \, .
\label{dc2/c2} \ee
Therefore, since for dark energy $w < 0 $ we see that $c^{2}$
augments with expansion. An additional consequence is that $c^{2}$
will vary sufficiently slow for $3\,|w|\, \Omega_{m} \lesssim 1$
only.

\bigskip

Let us continue by considering a holographic dark energy, whose
evolution is formally given by (\ref{conserv}), but choosing as
infrared cutoff the Ricci's length, $L= (\dot{H}+2 H^2)^{-1/2}$.
Thus,
\be \frac{c^2}{L^2}= \frac{8\pi G}{3}\, \rho_{x0}\,
 \exp \left[3 \int_{0}^{z}{\left(\frac{1+w(z')}{1+z'}
 \right)dz'}\right]\, ,
 \label{darkL} \ee
 i.e.,
 \be
 c^2= \frac{8\pi G}{3}\, \frac{\rho_{x0}}{\dot{H}+2H^2} \,
 \exp \left[3 \int_{0}^{t}{\left(1+w(t') \right) H(t') dt'}\right]\, .
 \label{darkRicci}
 \ee

From the field equations, and after some algebra, one can obtain
$c^2\left(1+r-3w\right) = 2$. This implies that
\be \frac{(c^{2})^{\cdot}}{c^{2}} =
\frac{\frac{\dot{w}}{w}-H r}{\frac{1}{3w\Omega_x}-1}.
\label{dc2/c2Ricci} \ee
Consequently, condition (\ref{condition}) amounts to
\be \frac{\dot{w}}{w} \lesssim H
\left[r-1+\frac{1}{3w\Omega_x}\right]. \label{wevol} \ee
In this case a constant $c^{2}$ is admissible provided the dark
energy density obeys \cite{gao}
\be \rho_{x}=\frac{\rho_{m0}c^2}{2-c^2}\left[a^{-3}+k
a^{-4+2/c^2}\right] \, , \label{constevol} \ee
where $k$ is an integration constant.

\begin{figure}[htb]
\centering
\includegraphics[width=9cm]{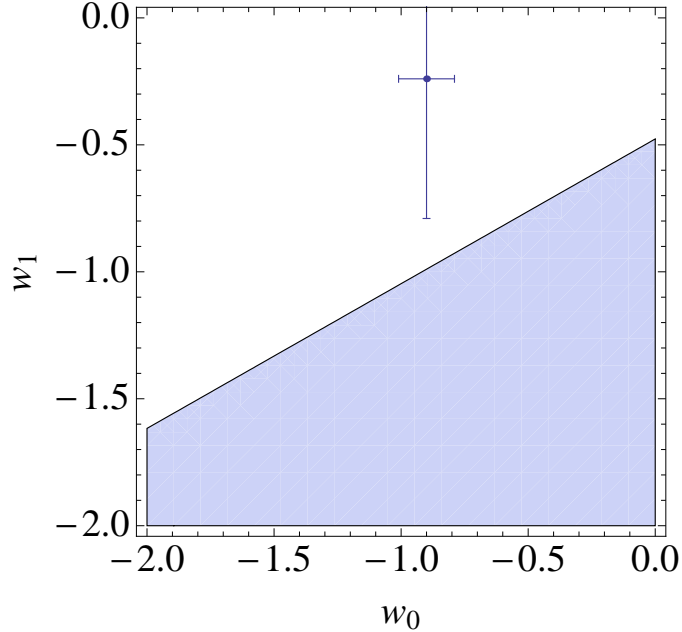}
\caption{The dark zone in the plane $(w_{0}, w_{1})$ corresponds
to the parameter space region in which the inequality
(\ref{wevol}), for $z=0$, is satisfied. Also indicated is the best
fit value with the $1\sigma$ uncertainties for the pair $(w_{0},
w_{1})$ as reported in \cite{zhao10}, -see Eqs. (\ref{zhaocon}).}
\label{fig:diseg}
\end{figure}

To check whether the bound (\ref{condition}) is fulfilled some
expression for $w$ is needed. We choose the widely used
Chevallier-Polarski-Linder parametrization \cite{chevallier01,
linder03},
\be w(z)=w_0+w_1\frac{z}{1+z} \, .
\label{varyingw} \ee
The constant parameters $w_{0}$ and $w_{1}$ are observationally
constrained by supernovae, cosmic background radiation, and large
scale structure data, see e.g. \cite{zhao10},
 \be w_0=-0.90^{+0.11}_{-0.11}\, , \quad
 \quad w_1=-0.24^{+0.56}_{-0.55}.
 \label{zhaocon}\ee
As inspection of Fig. \ref{fig:diseg} reveals, the holographic
constraint is fulfilled only at $2\sigma$ confidence level.

\bigskip

Finally, we repeat the analysis but now we choose the particle
horizon, defined in Eq. (\ref{parthor}), as infrared cutoff. In
this case,
\be c^{2}(a)= a^{2} f^{2}(a)
\left[\int_{0}^{a}\frac{da'}{\sqrt{r_0 a'+a'{^4}
f^2(a')}}\right]^{2}, \label{cvarpart} \ee
 where
 \be f(a)=\exp{\left[-\frac{3}{2}\, \int \frac{1+w(a')}{a'} da'\right]}
 \label{f(a)} \, .
 \ee
\noindent While Eq. (\ref{cvarpart}) does not completely exclude
the case of a constant $c^{2}$, this possibility looks rather slim
as $w(a)$ should take a very contrived expression.

To ascertain whether the bound (\ref{condition}) can be satisfied
we write the derivative of $c^2$ from the conservation equation
 \be
\frac{(c^{2})^{\cdot}}{c^{2}}=2H\left[1+\frac{\sqrt{\Omega_x}}{c}\right]-3H(1+w)
\label{cvarder1}
\ee
and use  $c/\sqrt{\Omega_x}=L H$ to get
\be \frac{(c^{2})^{\cdot}}{c^{2}}=H\left\{ 2
\left(1+\frac{1}{\sqrt{r_0 a^{-1}+a^2 f^2(a)} \int_{0}^{a}
\frac{da'}{a\sqrt{r_0 a'^{-1}+a'^{2} f^2(a')}}}\right)-3
(1+w)\right\}. \label{cvarder2} \ee
So, the term in the curly parenthesis should be lower than unity.
Even being rather conservative, inspection readily reveals that
for this to occur $w$ should be larger than $-2/3$ at all times,
which is widely excluded by observation.

\section{Concluding remarks}
Altogether, except for the particular case of dark energy with the
Ricci's length as infrared cutoff, in all the instances examined,
the $c^{2}$ term in the widely used holographic expression
(\ref{rhox}) for dark energy should not be assumed constant.
Generally, it varies faster than the Hubble rate at least for some
long periods of expansion. Even in the said particular case, using
the Chevallier-Polarski-Linder parametrization
\cite{chevallier01,linder03}, the bound (\ref{condition}) is
satisfied at just at $2\sigma$ confidence level.

Clearly the bound (\ref{condition}) looks reasonable but,
notwithstanding, it is debatable. On the one hand if $c^2$ varied
much more faster than the scale factor, then holography would
break down (i.e., the entropy of a region of size $L$ would not
longer be proportional to $L^{2}$). On the other hand, we do not
know of any definitive argument to adamantly enforce it. One may
contend that  $H$ could be replaced  by $n \, H$ on the right hand
side of (\ref{condition}), with $n$ a positive constant of order
unity. However, the precise value $n$ should take it is rather a
matter of choice. Therefore, for the sake of definiteness, we
believe it is better simply to keep $n = 1$.

We have not considered possible non-gravitational interactions in
the dark sector, very often invoked to alleviate the coincidence
problem -see eg. \cite{Amendola,srd2009,idz2010} and references
therein. It remains to be seen which specific interactions are
consistent with $c^{2} =$ constant or, more generally, with the
bound (\ref{condition}). We defer this study to a future work.

 \acknowledgments{We are grateful to Winfried Zimdahl for comments
on an earlier draft of this paper. NR is funded by the Spanish
Ministry of Education through the ``Subprograma Estancias de
J\'{o}venes Doctores Extranjeros, Modalidad B", Ref: SB2009-0056.
This research was partly supported by the Spanish Ministry of
Science and Innovation under Grant FIS2009-13370-C02-01, and the
``Direcci\'{o} de Recerca de la Generalitat" under Grant
2009SGR-00164.}


\end{document}